\documentclass[aps,a4paper,showpacs,superscriptaddress,twocolumn]{revtex4}

\usepackage{graphicx}
\usepackage{hyperref}
\usepackage[sort&compress]{natbib} % natbib for hyperrefed and sorted references; numbers: numerical citation; sort: sorted citations;
\usepackage[english]{babel}
\usepackage{mathrsfs}
\usepackage{amsmath}
\usepackage{amssymb}

\begin{document}
\title{On the analytic structure of scalar glueball operators at the Born level}
\date{\today}
\author{Andreas Windisch}
%\email[Electronic address: ]{andreas.windisch@uni-graz.at}
\affiliation{Institut f\"ur Theoretische Physik, Karl-Franzens-Universit\"at Graz, Universit\"atsplatz 5, 8010 Graz, Austria }
\author{Markus Q. Huber}
%\email[Electronic address: ]{markus.huber@uni-jena.de}
\affiliation{Institut f\"ur Kernphysik, Technische Universit\"at Darmstadt, Schlossgartenstrasse 2, 64289 Darmstadt, Germany}
\author{Reinhard Alkofer}
%\email[Electronic address: ]{reinhard.alkofer@uni-graz.at}
\affiliation{Institut f\"ur Theoretische Physik, Karl-Franzens-Universit\"at Graz, Universit\"atsplatz 5, 8010 Graz, Austria }

\begin{abstract}
We study the analytic structure of the two-point function of the operator $F^2$ which is expected to describe a scalar glueball. The calculation of the involved integrals is complicated by nonanalytic structures in the integrands, which we take into account properly by identifying cuts generated by angular integrals and deforming the contours for the radial integration accordingly. The obtained locations of the branch points agree with Cutkosky's cut rules. As input we use different nonperturbative Landau gauge gluon propagators with different analytic properties as obtained from lattice and functional calculations. All of them violate positivity and describe thus gluons absent from the asymptotic physical space. The resulting spectral densities for the glueball candidate show a cut but no poles for lightlike momenta, which can be attributed to the employed Born approximation.
\end{abstract}
\maketitle
%%%%%%%%%%%%%%%%%%%%%%%%%%%%%%%%%%%%%%%%%%%%%%%%%%%%%%%%%%%%%%%%%%%%%%%%%%%%%%%%%%%%%%%%%%%%%%%%%%%%%%%%%%%%%%%%%%%%%%%

\section{\label{intro}Introduction}

The Standard Model of particle physics describes all known elementary particles and their interactions. The sector of the strong nuclear force contains quarks and gluons, which, however, are not observed directly in experiments. Searches for partial electric charges (see, for example, \cite{Nash:1974dw,Antreasyan:1977va,Stevenson:1978wn,Bergsma:1984yn}) have all been negative and also gluons have not been observed \cite{HidalgoDuque:2011je}. This phenomenon is known as confinement and many scenarios trying to explain the underlying mechanism exist; for summaries see, for instance, \cite{Alkofer:2006fu,Greensite:2003bk}. What is observed in nature are bound states of quarks and gluons, which are said to be color neutral in contrast to their constituents that carry a color charge. Composite objects where the constituent degrees of freedom are only gluons are called glueballs. They are studied by a variety of different approaches; see, e.~g., \cite{Mathieu:2008me} for a review.

Regardless of any details of the confinement mechanism, one can study the spectral properties of quarks and gluons to look for violations of positivity \cite{Osterwalder:1973dx, Osterwalder:1974tc}. This is considered as a signal of confinement, since it means that such a particle has unphysical negative norm contributions and possesses no K\"all\'en-Lehmann representation \cite{Lehmann:1954xi,Kallen:1952zz}. Consequently such particles are absent from the physical subspace and are in this sense confined. For gluons positivity violations are established from lattice \cite{Langfeld:2001cz,Bowman:2007du} and functional calculations \cite{Alkofer:2003jj,Fischer:2008uz}. It is a nontrivial task to obtain physical glueballs from such positivity violating constituents. In order to construct meaningful operators the requirement to have a positive spectral density can serve as a guideline. In principle one can extract this information by studying the analytic structure of the two-point function of such an operator, which was done in several studies, see, for example, \cite{Zwanziger:1989mf, Baulieu:2009ha}.  Note that the spectral density can also be employed to extract glueball masses on the lattice \cite{Oliveira:2012eu}.

The search for bound states with functional approaches is usually performed by means of Bethe-Salpeter equations, which is a very successful approach for mesons and baryons. Only very recently it was extended to gluons \cite{Kellermann:2012th}. In the present study we take another approach and investigate the analytic structure of the two-point function of the squared Yang-Mills field strength tensor, the $F^2$ correlator. As a first step we only calculate the zeroth order in the coupling, i.~e., we work at the Born level. Thus no explicit self-interaction is considered. However, we account for gluonic interactions indirectly by using nonperturbative gluon propagators as an input which were obtained from the interacting theory. Obvious questions are whether the signature of a bound state shows up already at the Born level in this setting and if there are any major differences between the available gluon propagators.

If the gluon propagator input is simple enough, analytic calculations of the spectral representation are possible \cite{Zwanziger:1989mf,Baulieu:2009ha}. However, in general the integrals cannot be tackled by hand and a numerical approach is desirable. Furthermore, recently numerical solutions for the gluon propagator in the complex plane have become available \cite{Strauss:2012dg} which could also be used as an input to numeric calculations. We thus employ a numerical procedure  developed specifically for this purpose \cite{Windisch:2012zd}. The use of graphics processing units (GPUs) allows a high degree of parallelization and thus a high level of numerical precision.

Since no generally accepted analytic form of the gluon propagator is available, we study several different versions. They were obtained from fits to data obtained with lattice calculations or Dyson-Schwinger equations. Note that they stem from fits of data on the positive real axis only and we know from functional analysis that this is not enough to fix their analytic structure. First numerical results in the complex plane were obtained in \cite{Strauss:2012dg}. Some fits, however, are motivated by analytic results. The most prominent example is the tree-level expression of the Refined Gribov-Zwanziger (RGZ) action, which can be considered an effective action taking into account the restriction to the first Gribov region and the existence of several dimension two condensates \cite{Dudal:2008sp,Dudal:2011gd}. Already the restriction to the first Gribov region leads to complex conjugate poles of the tree-level gluon propagator \cite{Zwanziger:1989mf} and the condensates modify the gluon dressing function further. Since the suggested gluon propagators possess quite different analytic structures, it is interesting to see how they affect the structure of the $F^2$ correlator. We also study the two-dimensional case. Although not physically relevant, it allows us to test the method in a simplified setting.

In Sec. \ref{sec2} we provide the general expression of the $F^2$ correlator in our approximation. We investigate its analytic structure in Secs.~\ref{sec:res-2d} and \ref{sec:res-4d} for two and four dimensions, respectively, using different gluon propagators and conclude in Sec. \ref{summary}. A short summary of our findings can be found in Ref.~\cite{Windisch:2013mg}.

%%%%%%%%%%%%%%%%%%%%%%%%%%%%%%%%%%%%%%%%%%%%%%%%%%%%%%%%%%%%%%%%%%%%%%%%%%%%%%%%%%%%%%%%%%%%%%%%%%%%%%%%%%%%%%%%%%%%%%%
\section{\label{sec2}The correlator $\langle F^2(x)F^2(0)\rangle$  in $d$ Euclidean dimensions}

The quantity of interest is the correlator of the squared field strength tensor, which is a candidate for a scalar glueball:
\begin{equation}
\langle F^2(x)F^2(0)\rangle_d = \langle F_{\mu\nu}^a(x)F_{\mu\nu}^a(x)F_{\rho\sigma}^b(0)F_{\rho\sigma}^b(0)\rangle_d.
\label{eq1}
\end{equation}
$d$ is the (Euclidean) dimension. We will refer to this expression as the $F^2$ correlator.
At the Born level we neglect the non-Abelian part of the Yang-Mills field strength tensor:
\begin{equation}
F_{\mu\nu}^a=\partial_\mu A^a_\nu-\partial_\nu A^a_\mu.
\label{eq2}
\end{equation}
As customary we will investigate the momentum representation of the correlator:
\begin{equation}
\langle F^2(x)F^2(0)\rangle_d=\int\frac{d^dp}{(2\pi)^d}e^{i\,p\cdot x}\mathscr{O}_d(p^2),
\label{eq3}
\end{equation}
where $\mathscr{O}_d(p^2)$ reads \cite{Baulieu:2009ha}
\begin{widetext}
\begin{equation}
\mathscr{O}_d(p^2)=8C\int \frac{d^dk}{(2\pi)^d}\left(\mathscr{G}((p-k)^2)\mathscr{G}(k^2)(k^2(p-k)^2+(d-2)(k\cdot(p-k))^2)\right),
\label{eq4}
\end{equation} 
\end{widetext}
with $C=N_c^2-1$ and $N_c$ is the number of colors. In the following we will use $N_c=3$ for concreteness but it is only a trivial overall factor. Expression (\ref{eq4}) is valid for all kinds of gluon propagators which we parametrize as
\begin{equation}
D_{\mu\nu}(p^2)=\left(\delta_{\mu\nu}-\frac{p_\mu p_\nu}{p^2}\right)\mathscr{G}(p^2).
\label{eq5}
\end{equation}  
Note that in two dimensions the second term in the integrand vanishes.

In general the integral in Eq.~(\ref{eq4}) is UV divergent. To render it finite we employ BPHZ renormalization \cite{Bogoliubov:1957gp,Zimmermann:1969jj,Hepp:1966eg}. The highest degree of divergence that appears is of order $p^4$. Accordingly we subtract at most up to the fourth order in $p$:
\begin{align}
 &\mathscr{O}^r_d(p^2)=\mathscr{O}_d(p^2)\nonumber\\&-\mathscr{O}_d(0)-p^2\frac{\partial^2}{\partial p^2}\mathscr{O}_d(p^2)\Big|_{p=0}-p^4 \frac{\partial^4}{\partial  p^4}\mathscr{O}_d(p^2)\Big|_{p=0}.
\end{align}
Note that odd orders drop due to their asymmetric integrand. In the following we will drop the superscript $r$ again and always refer to renormalized quantities. Note that the operator $F^2$ can mix with other operators even in pure Yang-Mills theory \cite{Henneaux:1993jn,Dudal:2008tg}. However, such contributions are either BRST exact or proportional to the equations of motion. Thus they do not contribute to physical expectation values. This is no longer true for an action that breaks BRST invariance as the (Refined) Gribov-Zwanziger action; see \cite{Dudal:2009zh} for details.

The two-point function $\Delta(p^2)$ of a physical operator $\Phi$ (with zero spin) may be written in a spectral representation:
\begin{align}
 \Delta(p^2)=\int\frac{d^dp}{(2\pi)^d}e^{i\,p\cdot x} \langle \Phi(x) \Phi(0) \rangle=\int_{\tau_0}^\infty d\tau\frac{\rho(\tau)}{\tau+z},
\end{align}
where $\tau_0$ is the lowest possible energy of a state.
The analytic properties of $\Delta(p^2)$ are described entirely by this expression and can be read off from the spectral density $\rho(p^2)$. The spectral density $\rho(p^2)$ must be positive, which follows from positivity requirements in quantum theory. Violations of positivity are interpreted as an indication that such fields are absent from the physical spectrum, as found in the gluon propagator of Landau gauge Yang-Mills theory \cite{Langfeld:2001cz,Bowman:2007du,Alkofer:2003jj,Fischer:2008uz,Strauss:2012dg}.
The spectral density describes the jump across the branch cut along the timelike axis and can thus be calculated by
\begin{align}\label{eq:rho_disc}
 \rho(p^2)=\frac{1}{2\,\pi\,i}\lim_{\epsilon\rightarrow 0^+}[\Delta(-p^2-i\,\epsilon)-\Delta(-p^2+i\,\epsilon)]
\end{align}
with $-p^2>\tau_0$. Below we will calculate $\Delta(p^2)$ for the $F^2$ operator, from which we extract the spectral density via Eq.~(\ref{eq:rho_disc}).

%%%%%%%%%%%%%%%%%%%%%%%%%%%%%%%%%%%%%%%%%%%%%%%%%%%%%%%%%%%%%%%%%%%%%%%%%%%%%%%%%%%%%%%%%%%%%%%%%%%%%%%%%%%%%%%%%%%%%%%
\section{The analytic structure of the $F^2$ correlator in two dimensions}
\label{sec:res-2d}

We will first test the method in two dimensions. Two dimensions are advantageous because renormalization is easier: the integrand in Eq.~(\ref{eq4}) diverges quadratically in $p$. Furthermore, the second term drops. 
Two dimensions have also the advantage that only one type of solution exists \cite{Cucchieri:2012cb,Huber:2012zj,Zwanziger:2012xg}, whereas in four dimensions two qualitatively different kinds of solutions can be found; see Sec.~\ref{sec:res-4d}.
For the gluon propagator function $\mathscr{G}(k^2)$ we use the following fit form motivated originally for four dimensions in Ref.~\cite{Alkofer:2003jj}:
\begin{align}\label{eq:gl_2d}
 \mathscr{G}(p^2)=w\frac{1}{p^2}\left(\frac{p^2}{p^2+\Lambda^2}\right)^{1+2\kappa},
\end{align}
where $\kappa=0.2$ \cite{Zwanziger:2001kw}. Note that originally the possibility of $\kappa=0$ was also considered but can actually be excluded because it would require additional unphysical prescriptions to be a valid solution \cite{Huber:2012zj}. Equation~(\ref{eq:gl_2d}) describes the available data from lattice \cite{Maas:2007uv,Cucchieri:2007rg} and functional equations \cite{Huber:2012zj} in the IR for $w=1.065$ and $\Lambda=1\,\mathrm{GeV}$. Since deviations from the tree-level for high momenta go like $1/(1+c/p^2)$ the deviations in the UV are negligible. In the midmomentum regime very small deviations are observed, which are irrelevant for the present analysis. However, for easier comparison with $d=4$ we used here the same parameters as there, viz., $w=2.5$ and $\Lambda=0.51\,\mathrm{GeV}$.

\begin{figure}[tb]
\centering
\includegraphics[width=8cm]{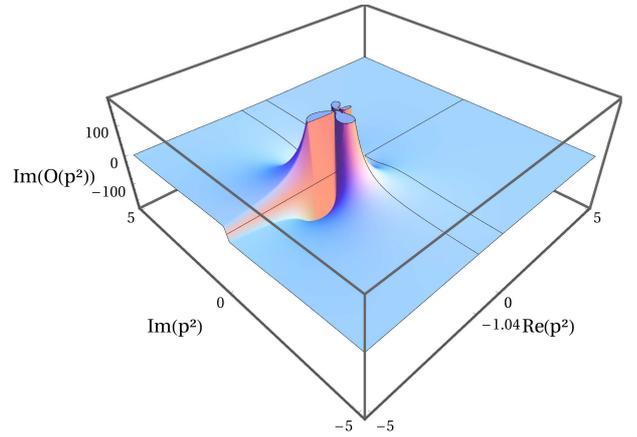}
\caption{The imaginary part of the $F^2$ correlator in 2d Landau gauge Yang-Mills theory with the scaling solution gluon propagator as input. A branch cut with a singular branch point appears on the negative real axis.}
\label{l2dim}
\end{figure}

\begin{figure}[tb]
\centering
\includegraphics[width=8cm]{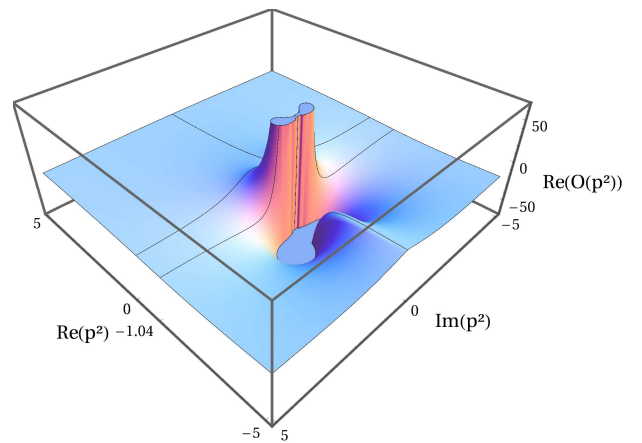}
\caption{The real part of the $F^2$ correlator in 2d Landau gauge Yang-Mills theory with the scaling solution gluon propagator as input.}
\label{l2dre}
\end{figure}

\begin{figure}[tb]
\centering
\includegraphics[width=8cm]{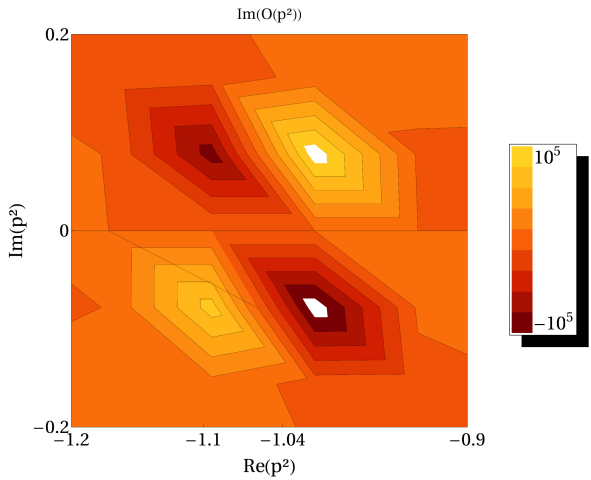}
\caption{A close-up contour plot of the pole structure of the singular branch point in the imaginary part of $\mathscr{O}(p^2)$.}
\label{l2dpole}
\end{figure}

\begin{figure}[tb]
\centering
\includegraphics[width=8cm]{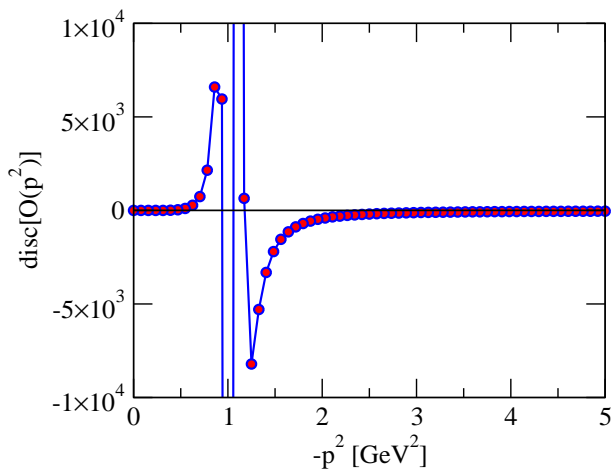}
\caption{The discontinuity of $\mathscr{O}(p^2)$ in 2d Landau gauge on the timelike axis is negative, thus no K\"all\'en-Lehmann representation exists.}
\label{l2ddisc}
\end{figure}

The results for $\mathscr{O}(k^2)$ are shown in Figs. \ref{l2dim} and \ref{l2dre}. On the timelike axis clearly a cut is observed. Since we consider this as a test case it is important to reproduce the expected position for the branch point. From a naive guess based on Cutkosky cut rules \cite{Cutkosky:1960sp}, which, however, are not directly applicable here (see, e.g., Ref.~\cite{Dudal:2010wn} for details), we expect the branch point at $p^2=-1.04\,\mathrm{GeV^2}$. Around this point a structure with four peaks appears, which are, however, of finite height as we checked explicitly. A close-up of this region is shown in Fig.~\ref{l2dpole}. The spectral density, as extracted via Eq.~(\ref{eq:rho_disc}), clearly turns negative, see Fig.~\ref{l2ddisc}. There one can also see some numerical artifacts, because the discontinuity already starts below $1.04$. However, for larger values, where the numeric evaluation is reliable, the spectral density is negative everywhere. Thus we conclude that $F^2$ is not a glueball. Of course, this is expected, since in two-dimensional Yang-Mills theory no transverse gluons exist which could compose a glueball and the spectrum must be trivial. Note that for the pure GZ propagator in two dimensions the spectral density is positive \cite{Baulieu:2009ha}.


%%%%%%%%%%%%%%%%%%%%%%%%%%%%%%%%%%%%%%%%%%%%%%%%%%%%%%%%%%%%%%%%%%%%%%%%%%%%%%%%%%%%%%%%%%%%%%%%%%%%%%%%%%%%%%%%%%%%%%%
\section{The analytic structure of the $F^2$ correlator in four dimensions}
\label{sec:res-4d}

In the four-dimensional case two qualitatively different types of gluon propagators can be found from functional equations depending on the boundary conditions chosen for the equations \cite{Fischer:2008uz}, the so-called decoupling \cite{Dudal:2008sp,Boucaud:2008ji,Aguilar:2008xm,Fischer:2008uz,Alkofer:2008jy} and scaling type solutions \cite{vonSmekal:1997is,vonSmekal:1997vx,Fischer:2008uz,Huber:2012kd}; see also the recent reviews \cite{Binosi:2009qm,Boucaud:2011ug,Maas:2011se}. The former has a nonvanishing value at zero momentum so that $\mathscr{G}(0)^{-1}$ can be interpreted as a screening mass, while the latter vanishes in the IR as $(p^2)^{2\kappa-1}$ with $\kappa=0.595353$ \cite{Zwanziger:2001kw,Lerche:2002ep}.

The tree-level propagator of the Gribov-Zwanziger action reads
\begin{align}
 \mathscr{G}(p^2)=\frac{p^2}{p^4+2\,N_c\,g^2\,\gamma^2},
\end{align}
where $\gamma$ is the so-called Gribov parameter. Qualitatively this propagator respects scaling with $\kappa=1$. This is true also for the ghost propagator, but then also the so-called horizon condition must be taken into account which renders it then divergent at one-loop level \cite{Zwanziger:1989mf}. This was confirmed also for two loops \cite{Gracey:2005cx}. Even more, in a nonperturbative treatment the same value for $\kappa$ emerges as for the standard Faddeev-Popov action \cite{Huber:2009tx,Huber:2010cq}. Thus the nonperturbative Gribov-Zwanziger propagators are contained in the class of scaling propagators investigated here, while the tree-level propagator is qualitatively the same but has a different analytic structure. The case of the Gribov-Zwanziger propagator can be worked out analytically \cite{Zwanziger:1989mf} and motivated the introduction of the so-called $i$-particles \cite{Baulieu:2009ha}, which allow the easy construction of an extension of $F^2$ from the Gribov-Zwanziger action that has only physical cuts. The analytic results of Ref.~\cite{Baulieu:2009ha} were used for testing the presently employed numeric code \cite{Windisch:2012zd}.

Within the Refined Gribov-Zwanziger framework the propagators change their qualitative behavior and become of the decoupling type \cite{Dudal:2008sp,Dudal:2007cw}. Here it will be important that in this case a certain analytic structure is determined at tree level. However, it is not clear if this structure is the one of the full nonperturbative propagator. The related case of the Gribov-Zwanziger action is one example where the difference between tree level and nonperturbative propagators is nicely illustrated \cite{Huber:2009tx,Huber:2010cq}.

For the scaling type we consider the nonperturbative part of the gluon propagator suggested in Ref.~\cite{Alkofer:2003jj} and for the decoupling type we investigate the propagator of the Refined Gribov-Zwanziger action \cite{Dudal:2008sp,Dudal:2011gd}. The latter only provides an effective description in the IR and midmomentum regimes, but the logarithmic running of the UV part is not expected to be relevant for the present analysis. Furthermore, the nonperturbative parts of other suggested fit forms, e.g., in Ref.~\cite{Aguilar:2007ie}, for the decoupling type solution have the same analytic structure with only minor changes in the parameters. Thus this choice covers a wide range of proposed fit forms. More about the specific choices of the fit forms will be said in the respective sections below.

%%%%%%%%%%%%%%%%%%%%%%%%%%%%%%%%%%%%%%%%%%%%%%%%%%%%%%%%%%%%%%%%%%%%%%%%%%%%%%%%%%%%%%%%%%%%%%%%%%%%%%%%%%%%%%%%%%%%%%%
\subsection{Decoupling propagator}
\label{sec:res-dec}

As explained above, one characterizing feature of the decoupling solution is a gluon propagator that becomes finite at zero momentum. Among possible fits with this property a propagator with a constant mass, i.e., 
\begin{align}\label{eq:prop-massive}
\mathscr{G}(p^2)\sim 1/(p^2+m^2),
\end{align}
cannot describe the available lattice data satisfactorily \cite{Bornyakov:2009ug,Dudal:2010tf}. A better description is provided by a momentum-dependent mass, i.~e., $m\rightarrow m(p^2)$. An often employed model for $m(p^2)$ is given by \cite{Aguilar:2007ie}
\begin{align}\label{eq:power-mass}
 m^2(p^2)=\frac{m_0^4}{p^2+m_0^2},
\end{align}
where we give only the IR relevant part. Terms taking into account the correct UV behavior can be found in \cite{Aguilar:2007ie}, where also the case of an IR constant mass term was investigated.
Within the RGZ framework the dimension two condensates raise the number of possible parameters, see Refs.~\cite{Gracey:2010cg,Dudal:2011gd} for several possibilities. A fit form motivated by the RGZ action that can successfully describe lattice data is given by \cite{Cucchieri:2011ig}
\begin{align}\label{eq:prop-RGZ}
\mathscr{G}(p^2)=C\frac{p^2+s}{p^4+u^2p^2+t^2}.
\end{align}
This propagator describes a more general case than the massive propagator in Eq.~(\ref{eq:prop-massive}) with the mass model from Eq.~(\ref{eq:power-mass}) and reduces to that for $s=u^2=t=m_0^2$. For the parameters we use the following values from Ref.~\cite{Cucchieri:2011ig}: $s=2.508\,\mathrm{GeV^2}$, $t=0.72\,\mathrm{GeV^2}$, $u=0.768\,\mathrm{GeV}$ and $C=0.784$. With these values the propagator has complex poles. Such a propagator form has also been used recently within a toy model including interactions \cite{Capri:2012hh}, where pole masses could be extracted using a bubble resummation.
We want to stress here that we take the fit form (\ref{eq:prop-RGZ}) provided by the RGZ action as an effective fit, but do not work with the Refined Gribov-Zwanziger action. This would entail an additional mixing of the operator $F^2$ with other operators \cite{Dudal:2009zh}.
 
\begin{figure}[tb]
\centering
\includegraphics[width=5cm]{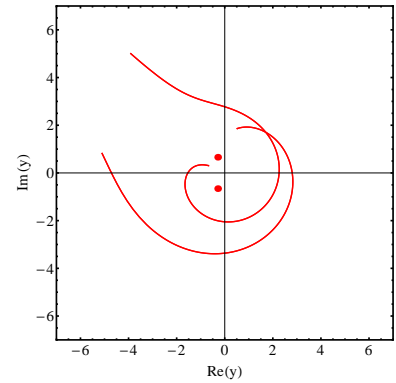}
\caption{The imaginary part of the analytically obtained result for the obstructive structure in the complex $y$ plane for $x=-2+2i$.}
\label{yplane-exact}
\end{figure}

\begin{figure}[tb]
\centering
\includegraphics[width=5.5cm]{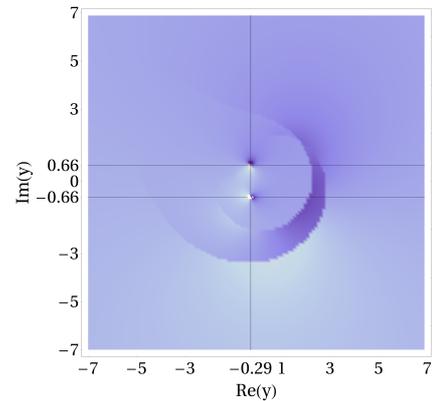}
\caption{The imaginary part of the radial integrand in the complex $y$ plane obtained from numerical integration for $x=-2+2i$. Besides the two branch cuts, a pair of complex conjugate poles is located at $p_{1,2}=-0.29\pm 0.66i$.}
\label{yplane}
\end{figure}

Before we proceed we want to make a few remarks concerning the analytic continuation of the momentum $x=p^2$ of the correlator to complex values. As soon as $x$ becomes a complex number, the square of the internal momentum $y=q^2$ has to be treated as a complex quantity as well. The reason for this is very simple: the generic structure of the correlators expressed in hyperspherical coordinates in four Euclidean dimensions is
\begin{align}\label{eq:generic}
\int_{0}^{\xi^2}dy y\int_{-1}^1dz\sqrt{1-z^2}f(x,y,z),
\end{align}
where $f(x,y,z)$ is the regularized integrand whose denominator can become singular for complex $x$. $z$ is the cosine of the angle between external and internal momenta. The outer radial integral in $y$ runs along the real axis from 0 to some cutoff in the UV denoted by $\xi^2$. As long as $x\in\mathbb{R}^+_0$, the contour for the radial integral exists and the integral can be evaluated. But if $x\notin\mathbb{R}^+_0$, branch cuts induced by the angular integral show up in the complex $y$ plane: as the angular integration variable $z$ runs through the interval of integration $[-1,1]$, it picks up a line of poles for singular values of $f(x,y,z)$.  In general, the contour along the real axis is obstructed by such a branch cut (sometimes more than one) which crosses the real axis on the interval ${[0,\xi^2]}$. The contour has to be deformed accordingly, in order to avoid these obstructive structures. In the case of the propagator given in Eq. (\ref{eq:prop-RGZ}), the induced obstructive structure in the complex $y$ plane is particularly complicated. It features a pair of complex conjugate poles as well as two branch cuts. The obstructive structure can be worked out analytically by finding the poles of the integrand for complex $x$ and $y$, where $z$ is used as a parameter to get the branch cuts. The results of the analytic calculation are compared with a numeric calculation for the (arbitrarily chosen) point $x=-2+2i$ in Figs. \ref{yplane-exact} and \ref{yplane}.


Thus, in addition to deforming the contour around the induced branch cuts, the poles have to be treated by either drawing the contours around the poles or by explicit subtraction of the residues after the radial integration. The latter approach of course also requires an integration around the pole in order to obtain the residue. We implemented both approaches and confirmed that they agree. However, each method has its strengths and weaknesses in certain regions; thus they can be applied according to their advantages. Note that the contour depends on the specific value of $x$, since the size and orientation of the cuts, but also the residues of the poles depend on it. Thus, in order to analytically continue the integrals to the complex $x$ plane, a thorough investigation of the complex $y$ plane after the angular integration is necessary. For more details on the whole procedure and a complete worked out example see Ref. \cite{Windisch:2012zd}.

\begin{figure}[tb]
\centering
\includegraphics[width=4.5cm]{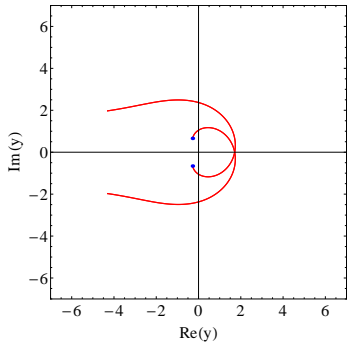}
\caption{At $x=-2$, the end points of the cuts in the $y$ plane are ''touching'' the poles. A valid contour connecting 0 to the UV cut-off $\xi^2$ \textit{must} in one way or another be drawn through one of these points. This restriction leads to a non-analyticity.
In fact, this point has also been obtained from a Cutkosky analysis.}
\label{obs1}
\end{figure}

\begin{figure}[tb]
\centering
\includegraphics[width=4.5cm]{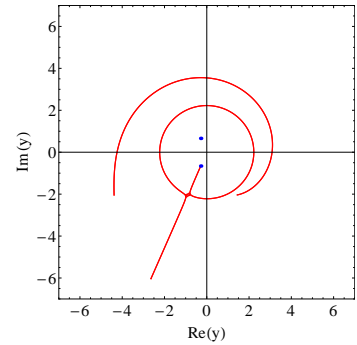}
\caption{At $x=-1.18-2.7i$ a valid contour must be drawn through the pinching curly cut, whose one end-point coincides with the pole. Again, the empirically determined value of $x$ where this happens is close to the Cutkosky prediction for the branch point.}
\label{obs2}
\end{figure}

\begin{figure}[tb]
\centering
\includegraphics[width=4.5cm]{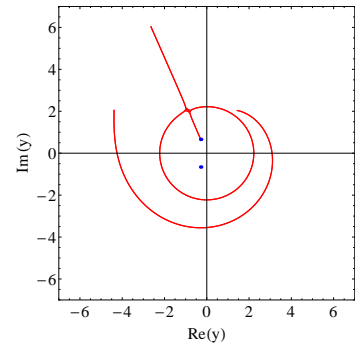}
\caption{For $x=-1.18+2.7i$ the situation is similar to the one found for the complex conjugate of $x$, as depicted in Fig.~\ref{obs2}.}
\label{obs3}
\end{figure}

\begin{figure}[tb]
\centering
\includegraphics[width=7cm]{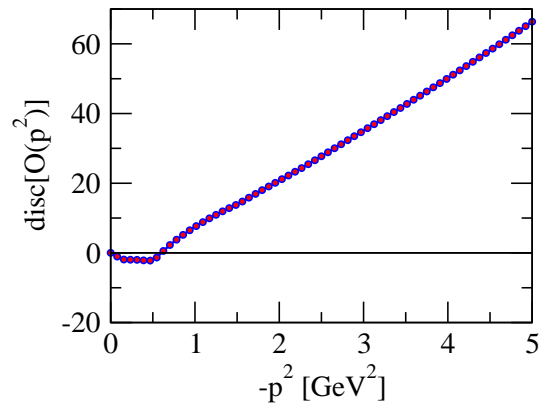}
\caption{The discontinuity of $\mathscr{O}(p^2)$ for the 4d Refined Gribov Zwanziger propagators. Some points are below the axis because of stability issues of the numerics. For the same reason the cut opens way too early as compared to the Cutkosky prediction. }
\label{rgz4ddisc}
\end{figure}

Further hints of what the resulting structure of the correlator in the $x$ plane might look like are provided by the Cutkosky cut rules. They allow us to calculate the positions of the expected branch points. For this purpose we need the two poles $p_1^2$ and $p_2^2$ of the gluon propagator. They are located at $p^2_{1,2}=-0.29\pm0.66\,i$. In order to apply the Cutkosky cut rules, we take these values to Minkowski space and calculate the position of the poles there from
\begin{align}
  \left(\sqrt{-p_i^2}+\sqrt{-p_j^2}\right)^2,
\end{align}
where $i$ and $j$ are $1$ or $2$. For $i\neq j$ we obtain a branch point at $2.03$. This leads to a cut in Euclidean space starting at $-2.03\,\mathrm{GeV^2}$. For $i=j$ the result is $1.18\pm i\, 2.63$; i.e., there are also two cuts in the complex plane starting at $-1.18\mp i\,2.63\,\mathrm{GeV^2}$. These cuts are considered unphysical in the sense that they forbid us to write a spectral representation for the correlation function of $F^2$. One motivation for the introduction of $i$-particles in Ref.~\cite{Baulieu:2009ha} was that they provided an easy way to get rid of such unphysical cuts. If these cuts were absent, we would have a proper spectral representation for the glueball, since the discontinuity across the cut on the timelike axis leads to a positive spectral density as shown in Fig.~\ref{rgz4ddisc}. The locations of the branch points as predicted from the Cutkosky analysis is also manifest in the obstructive structure in the complex $y$ plane. Upon varying the complex values of $x$ we could find three values of $x$ for which the obstructive structure in the $y$ plane is such that it leads to a nonanalyticity.
They are depicted in Figs. \ref{obs1}, \ref{obs2} and \ref{obs3}. The values were determined empirically and are in agreement with the three values obtained from the Cutkosky analysis. The empirical procedure of finding the pinching points gives rise to a small deviation from Cutkosky's predicted values. If they could be determined with arbitrary precision, they would coincide. Nontrivial analytic properties at these points are expected because in all three cases the integration contour \textit{must} be in one way or another drawn such, that it runs through the end point of one of the cuts which happens to lie just on top of one of the singularities. A possible path connecting 0 to the UV cutoff $\xi^2$ has to pass this point, which leads to a nonanalyticity.
Thus, the three points can also be found by looking for values of $x$ for which the possible contours have to be drawn exactly through the point where one (nonsingular) branch point ''touches'' one of the poles.


\begin{figure}[tb]
\centering
\includegraphics[width=7.5cm]{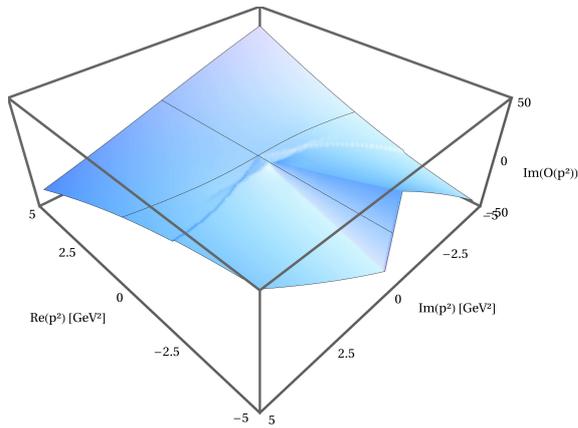}
\caption{The imaginary part of the $F^2$ correlator in 4d with Refined Gribov Zwanziger gluon propagators. The unphysical cuts open very slowly, making it almost impossible to determine the branch point location numerically.}
\label{rgz4dim}
\end{figure}

\begin{figure}[tb]
\centering
\includegraphics[width=7.5cm]{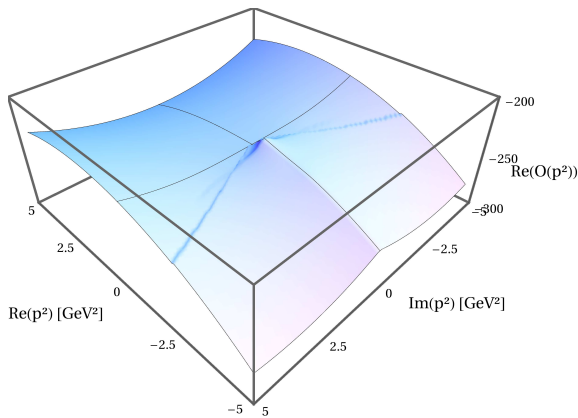}
\caption{The real part of the $F^2$ correlator in 4d with Refined Gribov Zwanziger gluon propagators.}
\label{rgz4dre}
\end{figure}

In Figs. \ref{rgz4dim}, \ref{rgz4dre} and \ref{rgz4ddisc} the numerical results for the imaginary part, the real part and the discontinuity along the physical cut of the correlator are shown, respectively. Because of the very restrictive analytic structure in the complex plane of the radial integration variable $y$, the exact structure is numerically almost not resolvable. Both methods described above give the same results in well-behaved regions, but tend to get unstable exactly in the directions of the three cuts. For the unphysical cuts this is simply due to the fact that the positions of the poles induced by the propagators have the same argument as the branch points. As the branch cuts always open into the direction given by $\arg(x)$, the only way the contour can be drawn out of the cuts is by going into the direction given by $\arg(x)$, i.e. running straight into the poles. All points having the same argument as the pole positions suffer from this problem. The unphysical cuts also have the property that they open very slowly, which makes a precise numerical determination of the branch point locations almost impossible. Nevertheless the small, but nonvanishing discontinuity clearly indicates the presence of the unphysical cuts. For small negative real numbers and the (almost) vanishing imaginary part, the contour runs very close to the obstructive branch cut over large distances, which makes it also very hard to obtain precise values in this region. Because of this, the physical branch cut opens too early as compared to the Cutkosky prediction. These stability issues can also be seen in the discontinuity of the physical cut, as shown in Fig. \ref{rgz4ddisc}. The few points below the axis can be attributed to the unstable numerics, thus the spectral function is strictly positive. The physical cut is clearly visible in fig. \ref{rgz4dim}. We want to stress that the results can be improved by adjusting the integration paths in the $y$ plane. However, this complicates the integration routine further and requires considerable fine-tuning and more integration path classes. To illustrate this we compare the effort required for the $i$-particle calculation of \cite{Windisch:2012zd} with that of the Refined Gribov-Zwanziger case: for the former, three different types of integration paths are necessary in order to avoid the singularities in the $y$ plane. Each of them consists of two to five integrations. For the Refined Gribov-Zwanziger propagator we employed here five different classes of paths. In order to obtain better results we expect that at least two more are necessary. Since it is not evident that the fit given in Eq.~(\ref{eq:prop-RGZ}) is the true nonperturbative form, as indicated also by numeric results for complex momenta \cite{Strauss:2012dg}, we refrain from this further complication. Nevertheless, we could demonstrate how the cuts in this case arise and obtained results for the branch point positions in agreement with the Cutkosky cut rules.
%%%%%%%%%%%%%%%%%%%%%%%%%%%%%%%%%%%%%%%%%%%%%%%%%%%%%%%%%%%%%%%%%%%%%%%%%%%%%%%%%%%%%%%%%%%%%%%%%%%%%%%%%%%%%%%%%%%%%%%

%%%%%%%%%%%%%%%%%%%%%%%%%%%%%%%%%%%%%%%%%%%%%%%%%%%%%%%%%%%%%%%%%%%%%%%%%%%%%%%%%%%%%%%%%%%%%%%%%%%%%%%%%%%%%%%%%%%%%%%
\subsection{Scaling propagator}
\label{sec:res-scal}

The nonperturbative part of the four-dimensional scaling gluon propagator can be described by \cite{Alkofer:2003jj}
\begin{align}\label{eq:prop-scal}
 \mathscr{G}(p^2)=w\frac{1}{p^2}\left(\frac{p^2}{p^2+\Lambda^2}\right)^{2\kappa},
\end{align}
where $\kappa=0.595353$ \cite{Zwanziger:2001kw,Lerche:2002ep}. In order to account for the full momentum dependence of the gluon propagator, this function has to be multiplied by another function that describes the UV logarithmic tail \cite{Alkofer:2003jj}. However, we are here only interested in the nonperturbative content of the propagator and thus use expression (\ref{eq:prop-scal}). This also avoids technical complications related to the deformation of the contour in the integration process. For the parameters we use the value from the fit in \cite{Alkofer:2003jj}, viz. $\Lambda=0.51\,\mathrm{GeV}$ and $w=2.5$.

\begin{figure}[tb]
\centering
\includegraphics[width=7.5cm]{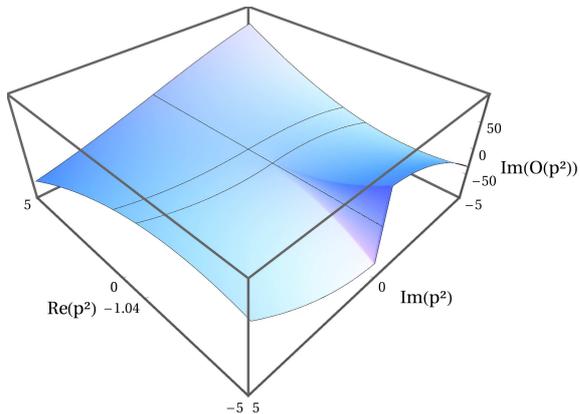}
\caption{The imaginary part of the $F^2$ correlator in 4d Landau gauge Yang-Mills theory with scaling solution gluons as input.}
\label{l4dim}
\end{figure}

\begin{figure}[tb]
\centering
\includegraphics[width=7.5cm]{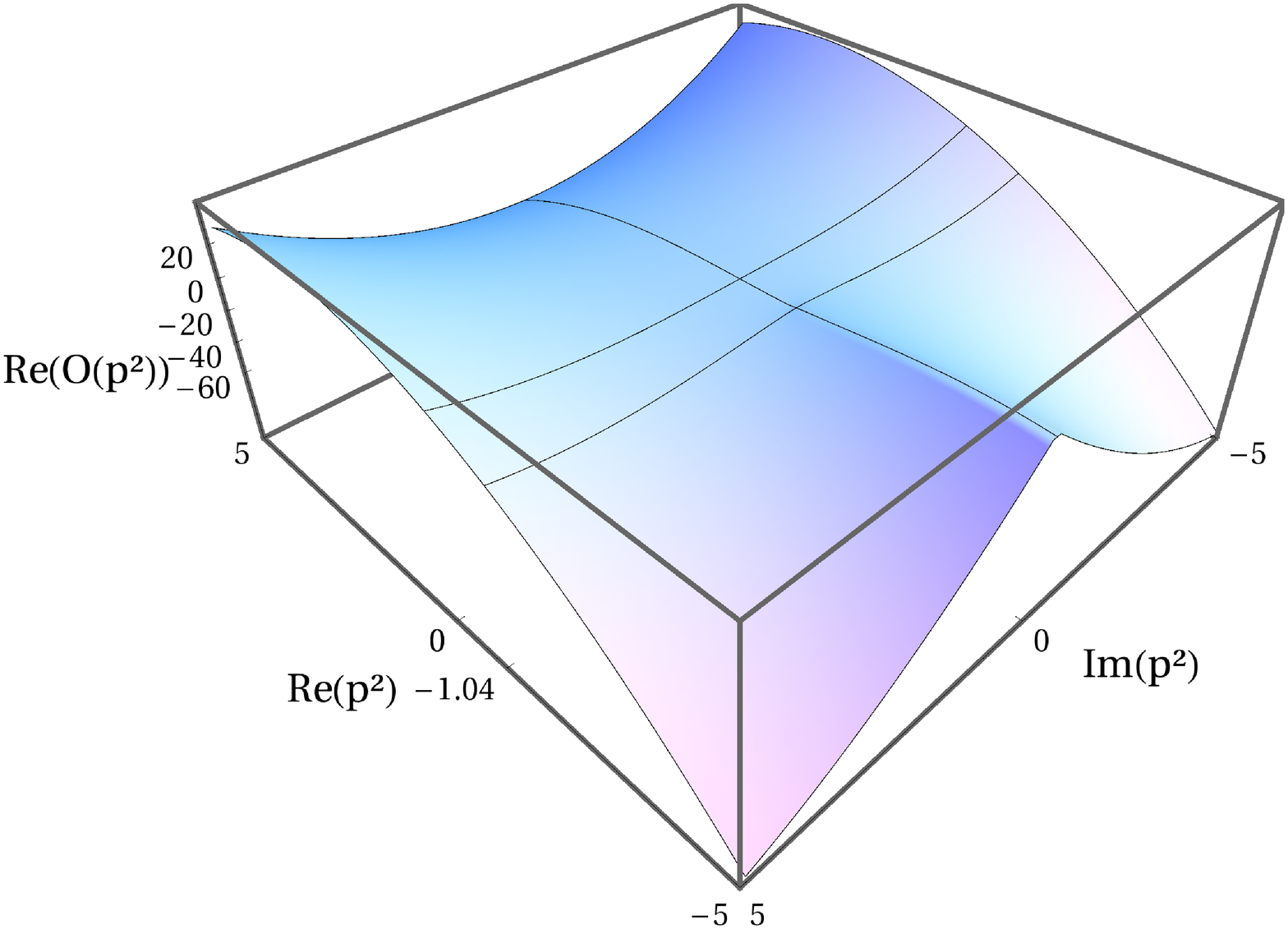}
\caption{The real part of the $F^2$ correlator in 4d Landau gauge Yang-Mills theory with scaling solution gluons as input.}
\label{l4dre}
\end{figure}

\begin{figure}[tb]
\centering
\vspace{0.2cm}
\includegraphics[width=7cm]{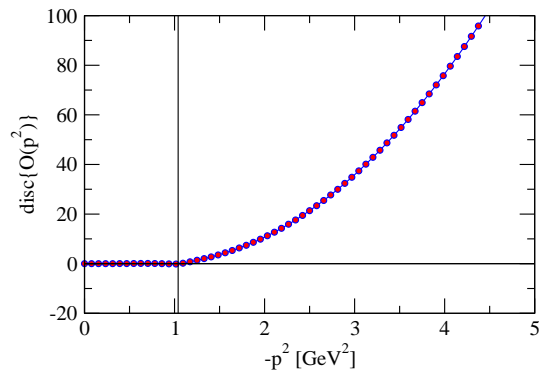}
\caption{The discontinuity of $\mathscr{O}(p^2)$ using scaling solution gluon propagators.}
\label{l4ddisc}
\end{figure}

The resulting correlation function of $F^2$ is shown in Figs.~\ref{l4dim} and \ref{l4dre}. As expected, there is a branch cut on the timelike axis; see Fig.~\ref{l4dim}. The jump of the correlation function, which is directly related to its spectral density, is plotted in fig.~\ref{l4ddisc}. In this case the numerics are perfectly under control and we can easily deduce the position of the branch point, which is $1.04\,\mathrm{GeV^2}$. We thus confirm that the $F^2$ operator corresponds to a physical quantity with a positive spectral density. However, we do not observe any poles which, of course, are not expected to appear at Born level.

%%%%%%%%%%%%%%%%%%%%%%%%%%%%%%%%%%%%%%%%%%%%%%%%%%%%%%%%%%%%%%%%%%%%%%%%%%%%%%%%%%%%%%%%%%%%%%%%%%%%%%%%%%%%%%%%%%%%%%%

%%%%%%%%%%%%%%%%%%%%%%%%%%%%%%%%%%%%%%%%%%%%%%%%%%%%%%%%%%%%%%%%%%%%%%%%%%%%%%%%%%%%%%%%%%%%%%%%%%%%%%%%%%%%%%%%%%%%%%%
\section{\label{summary}Summary and conclusions}
%%%%%%%%%%%%%%%%%%%%%%%%%%%%%%%%%%%%%%%%%%%%%%%%%%%%%%%%%%%%%%%%%%%%%%%%%%%%%%%%%%%%%%%%%%%%%%%%%%%%%%%%%%%%%%%%%%%%%%%

In this work we investigated the analytic properties of the correlation function of the operator $F^2$, which is a candidate for a scalar glueball. In our setup we used different nonperturbative gluon propagators which are obtained from fits to lattice and/or continuum results. For some simple cases the calculations can be done analytically. However, for future investigations numerical methods are required for several reasons. First of all, one would like to directly use results for the gluon propagator. Such data became available only recently \cite{Strauss:2012dg}. In this case no fits are available and numeric data are the input. Second, extensions of the current scheme most likely can only be calculated numerically. With the present work we have laid the basis for such calculations. In any case it will be useful to continue as done here with GPUs as they allow a great deal of parallelization.

Since we worked at the Born level, we did not find any poles but only cuts for the $F^2$ correlator. 
The corresponding branch points on the real axis constitute the multiparticle threshold and thus an upper bound on glueball masses. The extracted values are $1.42\,\mathrm{GeV}$ and $1.02\,\mathrm{GeV}$ for decoupling and scaling, respectively. These values strongly depend on the parameter fits. However, in the scaling case the parameter $\Lambda$, which directly sets the scale, is not determined uniquely, since we considered only the low momentum part of the fit. Furthermore, it was obtained from a fit to DSE results which show a gap in the midmomentum regime \cite{Alkofer:2003jj}. Taking this into account will influence the value of $\Lambda$ and thus move also the threshold. Due to these uncertainties it looks more promising to employ numerical data obtained for complex momenta as input in the future as discussed above. For the decoupling type propagator two additional cuts in the complex plane were found that forbid a spectral representation. However, at this stage we cannot draw any physical conclusions from this difference between the two solutions, because the unphysical cuts can directly be attributed to the analytic properties of the decoupling fit. This shows that fits of the class described in Eq.~(\ref{eq:prop-RGZ}) are most likely inadequate to describe the analytic structure of the gluon propagator. While the employed fit form can describe lattice data reasonably well, this seems no longer to be true for complex momenta. Indications for this are also found by a direct calculation of the gluon propagator \cite{Strauss:2012dg}, where no poles in the complex plane have been found. 

Several possibilities for future calculations exist. On the technical side we have the inclusion of interactions, while on the formal level a better input is required for the gluon propagators. Due to recent results \cite{Strauss:2012dg} this may even come from calculations directly for complex momenta instead of fits to Euclidean momenta. The work presented here will in any case provide useful guidance for determining the analytic structure of correlators.

\section{Acknowledgments}
We thank David Dudal and Lorenz von Smekal for useful discussions.
MQH was supported by the Alexander von Humboldt foundation. AW acknowledges support by the Doktoratskolleg ``Hadrons in Vacuum, Nuclei and Stars'' funded by the FWF under Contract No. W1203-N16.

\bibliographystyle{utphys_mod}
\bibliography{literature_F2}

\end{document}